\documentclass{article} 
\usepackage[all]{xy} 
\usepackage{amsfonts} 
\usepackage{amsmath} 
\usepackage{eucal} 
\usepackage{amssymb}

\newcommand{\eq}{\begin{equation}} 
\newcommand{\feq}{\end{equation}} 
\newcommand{\eqn}{\begin{eqnarray}} 
\newcommand{\feqn}{\end{eqnarray}} 
\newcommand{\arr}{\begin{eqnarray*}} 
\newcommand{\farr}{\end{eqnarray*}}

\begin{document}

\begin{titlepage} 
\begin{flushright} 
IFUM--846--FT\\ 
LBNL--57265\\ 
UCB--PTH--05/05 
\end{flushright} 
\vskip 10mm 
\begin{center} 
\renewcommand{\thefootnote}{\fnsymbol{footnote}} 
{\Large \bf On the Euler angles for \boldmath{$SU(N)$}} 
\vskip 12mm 
{\large \bf {Stefano Bertini$^2$\footnote{bertiste@tiscalinet.it} 
Sergio L.~Cacciatori$^{1,3}$\footnote{sergio.cacciatori@mi.infn.it} and 
Bianca L. Cerchiai$^{4,5}$\footnote{BLCerchiai@lbl.gov}}}\\ 
\renewcommand{\thefootnote}{\arabic{footnote}} 
\setcounter{footnote}{0} 
\vskip 5mm 
{\small 
$^1$ Dipartimento di Matematica dell'Universit\`a di Milano,\\ 
Via Saldini 50, I-20133 Milano, Italy. \\ 
 
\vspace*{0.4cm} 
 
$^2$ Dipartimento di Fisica dell'Universit\`a di Milano,\\ 
Via Celoria 16, 
I-20133 Milano, Italy.\\ 
 
\vspace*{0.4cm} 
 
$^3$ INFN, Sezione di Milano,\\ 
Via Celoria 16, 
I-20133 Milano, Italy.\\ 
 
\vspace*{0.4cm} 
 
$^4$ Lawrence Berkeley National Laboratory\\ 
Theory Group, Bldg 50A5104\\ 
1 Cyclotron Rd, Berkeley, CA 94720-8162, USA.\\ 
 
\vspace*{0.4cm} 
 
$^5$ Department of Physics, University of California,\\ 
Berkeley, CA 94720, USA. 
} 
\end{center} 
\vspace{1cm} 
\begin{center} 
{\bf Abstract} 
\end{center}      
{\small  In this paper we reconsider the problem of the Euler parametrization 
for the unitary groups. After constructing the generic group element in 
terms of generalized angles, we compute the invariant measure on $SU(N)$ 
and then we determine the full range of the parameters, using both topological 
and geometrical methods. In particular, we show that the given 
parametrization realizes the group $SU(N+1)$ as a fibration of $U(N)$ over 
the complex projective space $\mathbb{CP}^n$. This justifies the  
interpretation of the parameters as generalized Euler angles.} 
 
\end{titlepage} 
 
 
\section{Introduction} 
The importance of group theory in all branches of physics is a well-known fact. 
Explicit realizations of group representations are often necessary technical 
tools. Often it is finite dimensional and compact Lie groups and then  
the knowledge of the associated algebra, which describes the group in a 
neighborhood of the identity, is enough for this purpose.\\ 
There are however cases where an explicit expression of the full global 
group structure is needed, as for example when non perturbative 
computations come into play. 
In most of these cases, the main objectives are two: First, one would like 
to find a relative simple parametrization, making all the computations 
manageable. Second, one needs to determine the full range of the parameters, 
in order to be able to handle global questions.\\ 
If both such points can seem unnecessary at an abstract level, they become 
essential at a most concrete level, e.g. in instantonic calculus 
or in nonperturbative lattice gauge theory computations. The necessary 
computer memory for simulations is in fact drastically diminished.\\ 
The case of $SU(N)$ was first considered and solved by Tilma and Sudarshan, 
in \cite{Tilma}. There, they provide a parametrization, in terms of angular 
parameters, for the unitary groups. In particular, in the first paper they 
consider special groups, $SU(N)$, together with some applications to qubit and 
qutrit configurations. In the second paper, they give an extension 
to $U(N)$ groups, using the fibration structure of $SU(N+1)$ as $U(N)$  
fiber over the complex projective space $\mathbb{CP}^n$.\\ 
In this paper we reconsider the problem of finding a generalized Euler 
parametrization for special unitary groups. The intent is to provide a fully 
explicit and elementary\footnote{which doesn't means short!} proof of the  
beautiful results of \cite{Tilma}. Our motivation is that the determination 
of the range of the parameters is a quite difficult task, 
so that disagreements are present in the literature even for $SU(3)$ 
(for example in \cite{Cvetic:2001zx}). Therefore, we think that a careful 
deduction is necessary in order to corroborate the results of Tilma and 
Sudarshan. Also, all our proofs based essentially on inductive procedures, 
and they are explicit, in order to be easily accessible to anyone who  
needs them.\\ 
Our construction is quite different from \cite{Tilma}, and 
as a result our parametrization differs slightly from theirs.  
However, this doesn't affect the final expression of the invariant measure. 
 
To illustrate the spirit of our construction, let us start by taking a look 
at the Euler parametrization for $SU(2)$.\\ 
Starting from the Pauli matrices 
\eqn 
\sigma_1 =\left( \begin{array}{cc} 0 & 1 \\ 1 & 0 \end{array} \right) \ , 
\qquad 
\sigma_2 =\left( \begin{array}{cc} 0 & -i \\ i & 0 \end{array} \right)\ ,\qquad 
\sigma_3 =\left( \begin{array}{cc} 1 & 0 \\ 0 & -1 \end{array} \right) \ , 
\feqn 
it is known that the generic element of $SU(2)$ can be written as 
\eqn 
g= e^{i\phi \sigma_3} e^{i\theta \sigma_2} e^{i\psi \sigma_3}  \ .  \label{su2} 
\feqn 
Here $\phi \in [0, \pi]$, $\theta \in [0, \pi/2]$, $\psi \in [0,2\pi]$ are 
the so called Euler angles for $SU(2)$. They are related to the well-known  
Euler angles traditionally used in classical mechanics to describe the motion 
of a spin. 
{From} the point of view of the structure of the representation, 
(\ref{su2}) is obtained starting from a one parameter subgroup  
$exp (i\theta \sigma_2)$ and then acting on it both from the left and from  
the right with a maximal subgroup of $SU(2)$ which doesn't contain the first 
subgroup. We can rewrite it in the schematic form $g= U(1)exp (i\theta 
\sigma_2) U(1)ÿ$. On the other hand, the group $SU(2)$ is topologically 
equivalent to the three-sphere $S^3$, and admits a Hopf fibration structure 
with fiber $S^1$ over the base $S^2 \simeq \mathbb{CP}^1$.\\ 
To recognize this fibration structure in (\ref{su2}), we can apply the 
methods used in \cite{Cacciatori:2005yb,Cacciatori:2005gi}. 
After introducing the metric $\langle A |B \rangle =\frac 12 Tr (AB)$ on the 
algebra, the metric on the group can be computed as $ds^2 =J \otimes J$, 
where $J =-ig^{-1} dg$ are the left-invariant currents. 
Following \cite{Cacciatori:2005yb}, it is possible to separate the fiber 
from the base by writing $g=h U(1)$, where 
$h=e^{i\phi \sigma_3} e^{i\theta \sigma_2}$ 
and $U(1)=e^{i\psi \sigma_3}$. To find the metric on the fiber, let's  
fix the point on the base and compute the currents along the fiber, 
$J_F =-iU(1)^{-1}dU(1) =d\psi \sigma_3$. The metric on the fiber is then simply 
given by $ds_F^2 = d\psi^2$. \\ 
To determine the metric on the base, we first have to project out from 
the current $J_B =-i h^{-1} dh$ the component along the fiber, in order to be 
left with the reduced current on the basis 
$\tilde J_B =d\psi \sigma_2 +\sin (2\psi) d\psi \sigma_1$, which then in turn  
provides the metric 
\eqn 
ds^2_B = \frac 14 \left[ d(2\psi)^2 +\sin^2 (2\psi) d(2\phi)^2 \right] \ . 
\feqn 
This corresponds in fact with the metric of a sphere of radius $\frac 12$. 
It is easy to see that, introducing the complex coordinates 
$z=\tan \psi e^{i\phi}$ and their complex conjugates, the metric $ds^2_B$ 
reduces to the standard Fubini-Study metric for $\mathbb{CP}^1$. \\ 
This shows that the Euler parametrization captures the Hopf fibration 
structure of $SU(2)$, which is the starting point for our construction. 
Mimicking what we said about $SU(2)$, let's write the generic element of 
$SU(N+1)$ as $g=U(N) e(\theta) U(N)$, where $e(\theta)$  is a one parameter 
subgroup not contained in $U(N)$. The first difficulty we have to face here 
is that this expression for a generic $SU(N+1)$ group has redundancies, 
which have to be eliminated. After this problem is solved, we then have 
to show that the parametrization respects the Hopf fibration structure 
of $SU(N+1)$.

 
\section{The \boldmath{$SU(N)$} algebra} 
\label{algebra} 
\setcounter{equation}{0} 
The generators of $su(N)$ are all the $N\times N$ traceless hermitian matrices. 
A convenient choice for a base are the generalized Gell-Mann matrices as 
explained in \cite{Tilma}. Let's remind how they can be constructed 
using an inductive procedure. Let be $\{\lambda_i\}_{i=1}^{N^2-1}$ the 
Gell-Mann base for $su(N)$: They are $N\times N$ matrices which can be 
embedded in $su(N+1)$ adding a null column and a null row 
\eqn 
\tilde \lambda_i = 
\left( 
\begin{array}{cc} 
\lambda_i & \vec 0 \\ 
\vec 0 & 0 
\end{array} 
\right) \ . 
\feqn 
We will omit the tilde from now on. The dimension of $SU(N)$ being $(N+1)^2-1$, 
we must add $2N+1$ matrices to obtain a Gell-Mann base for $su(N+1)$. This can 
be done as follows: Put 
\eqn 
&&\{\lambda_{N^2+2a-2}\}_{\alpha \beta} =\delta_{\alpha ,a} \delta_{\beta ,N+1} 
+\delta_{\alpha ,N+1} \delta_{\beta ,a} \ , \cr 
&&\{\lambda_{N^2+2a-1}\}_{\alpha \beta} =i(-\delta_{\alpha ,a} \delta_{\beta ,N+1} 
+\delta_{\alpha ,N+1} \delta_{\beta ,a}) \ , 
\feqn 
for $a=1,\ldots , N$. The last matrix we need is diagonal and traceless so that 
we can take $\lambda_{(N+1)^2-1}= \epsilon_{N+1} diag \{ 1,\ldots, 1, -N \}$.\\ 
One can easily verify that the base of matrices  
$\{ \lambda_I \}_{I=1}^{(N+1)^2-1}$ 
so obtained satisfies the normalization condition 
$Trace \{ \lambda_I \lambda_J \} = 
2\delta_{IJ}$ if we choose $\epsilon_{N+1} =\sqrt {\frac 2{(N+1)^2-(N+1)}}$.\\ 
These are exactly the matrices we need to generate the group elements. 
 
 
\section{The Euler parametrization for 
\boldmath{$SU(N+1)$}: Inductive construction} 
 
\setcounter{equation}{0} 
It is a well known fact that special unitary groups $SU(N+1)$ can be 
geometrically understood as $U(N)$ fibration over the complex projective 
space $\mathbb{CP}^N$. Now $U(N)$ is generated 
by the first $N^2-1$ generalized Gell-Mann matrices plus the last one 
$\lambda_{(N+1)^2-1}$. Using the fact that all the remaining generators of 
$SU(N+1)$ can be obtained from the commutators of these matrices  
with $\lambda_{N^2+1}$, one is tempted to write the general element of 
$SU(N+1)$ in the form 
\eqn 
SU(N+1) =U(N) e^{ix \lambda_{N^2+1}} U(N) \ . 
\feqn 
However to describe $SU(N+1)$ we need $(N+1)^2-1$ parameters, while in the 
r.h.s. they are $2N^2 +1$: There are $(N-1)^2$ redundancies. Inspired at 
first by dimensional arguments, we propose that an $U(N-1)$ subgroup can be 
subtracted from the left $U(N)$ in the following way.\\ 
Let us write $U(N)$ in the form $U(N)=SU(N) e^{i\psi \lambda_{(N+1)^2-1}}$. 
Inductively, we can think that also $SU(N)$ can be recovered from  
$U(N-1) e^{i\phi \lambda_{(N-1)^2+1}} U(N-1)$ eliminating the redundant 
parameters, so that it will have the form 
$SU(N) =h e^{i\phi \lambda_{(N-1)^2+1}} SU(N-1) e^{i\theta \lambda_{N^2-1}}$. 
We then choose to eliminate the appearing $SU(N-1)$ together with the phase 
$e^{i\psi \lambda_{(N+1)^2-1}}$. In this way the $SU(N+1)$ group element can 
be written in the form $SU(N+1)=h e^{i\phi \lambda_{(N-1)^2+1}} 
e^{i\theta \lambda_{N^2-1}} e^{ix \lambda_{N^2+1}} U(N)$. 
By induction, assuming $N\geq 2$ we arrive to the final form of our Ansatz 
about the parametrization of the general element $g\in SU(N+1)$ 
\eqn 
g=e^{i\theta_1 \lambda_3} e^{i\phi_1\lambda_2} \prod_{a=2}^N [e^{i  
\frac{\theta_a}{\epsilon_a} \lambda_{a^2-1}} e^{i\phi_a\lambda_{a^2+1}} ]  
U(N)[\alpha_1,\ldots , \alpha_{N^2}] \ , \label{ansatz} 
\feqn 
where 
$U(N)[\alpha_1,\ldots , \alpha_{N^2}]$ is a parametrization of $U(N)$ which 
in turn can be obtained inductively using the fact 
\eqn 
U(N) =[SU(N)\times U(1)]/\mathbb{Z}_N \ . \label{uenne} 
\feqn 
The Ansatz (\ref{ansatz}) contains the correct number of parameters. 
However, we need to show that it is a good Ansatz, meaning that at least 
locally it has to generate the whole tangent space to the identity. 
Using the Backer-Campbell-Hausdorff formula and some properties of the 
Gell-Mann matrices\footnote{Essentially the fact that the commutators of 
$\lambda_{(k-1)^2+1}$ with the first $(k-1)^2-1$ matrices generate all the  
remaining matrices of the $su(k)$ algebra but the last one},  
it is easy to show that 
\eqn 
e^{i\theta_1 \lambda_1} e^{i\phi_1\lambda_3} \prod_{a=2}^N [e^{i  
\frac {\theta_a}{\epsilon_a} \lambda_{a^2-1}}  
e^{i\phi_a\lambda_{a^2+1}} ] =e^{i \sum_{j=1}^{(N+1)^2-2} a_j \lambda_j} \ , 
\feqn 
where $a_j$ are all non vanishing functions of the $2N$ parameters 
$\theta_a , \phi_a$. Thus in a change of coordinates  
(from the $\theta_a , \phi_a$ to the $a_j$) only $2N$ of the $a_j$ can be 
chosen as independent parameters. We could choose the last ones, 
corresponding to the coefficients of the matrices 
$\{ \lambda_k  \}_{k=N^2}^{N^2+2N-1}$. In this way, the $N^2$ 
free parameters for the remaining matrices come out exactly from the 
$U(N)$ factors in (\ref{ansatz}).\\ 
We have not entered into details here because a second simple proof of the 
validity of this parametrization will be given by constructing a nonsingular 
invariant measure from our Ansatz. 
 
 
\section{Invariant measure and the range of the parameters} 
\setcounter{equation}{0} 
\subsection{The invariant measure} 
To construct the invariant measure for the group starting from (\ref{ansatz}), 
we will adopt the same method used in 
\cite{Cacciatori:2005yb}, with $U :=U(N)$ as the fiber group. Let us then 
write (\ref{ansatz}) as 
\eqn 
g=h \cdot U \ . 
\feqn 
Starting from the computation of the left invariant currents 
$j_h =-ih^{-1} dh$, we can define the one forms 
\eqn 
e^l := \frac 12 Tr \ [ j_h \cdot \lambda_{N^2+l-1} ] \ , \qquad l=1,\ldots,2N  
\ , \label{vielbein} 
\feqn 
which turns out to give the Vielbein one forms of the base space of the 
fibration. 
If $\underline e$ denotes the corresponding Vielbein matrix, the invariant 
measure for $SU(N+1)$ will then take the form 
\eqn 
d\mu_{SU(N+1)} =\det \underline e \cdot d\mu_{U(N)} \ , 
\feqn 
$d\mu_{U(N)}$ being the invariant measure for $U(N)$. Using (\ref{uenne}) with  
$U(1) =e^{i \frac \omega{\epsilon_{N+1}} \lambda_{(N+1)^2-1} }$  we obtain the 
recursion relation\footnote{Note that here $\omega$ is allowed to vary in the  
range $[0,2\pi/N]$.} 
\eqn 
d\mu_{SU(N+1)} =\det \underline e \cdot d\mu_{SU(N)}  
\frac {d\omega}{\epsilon_{N+1}} \ . 
\feqn 
Then we will concentrate on the $\det \underline e$ term. 
To this end let us write (\ref{ansatz}) in the form 
\eqn 
g=h_{N+1} [\theta_a , \phi_a] \cdot U[\alpha_i] \ . 
\feqn 
Here we will consider $N\geq 3$ so that the relation 
\eqn 
h_{N+1} =h_N e^{i\frac {\theta_N}{\epsilon_N}\lambda_{N^2-1}} e^{i\phi_N 
\lambda_{N^2+1}} \ , 
\feqn 
is true. If we introduce the right currents $J_{h_{N+1}} =-i h_{N+1}^{-1}  
dh_{N+1}$ 
then the Vielbein (\ref{vielbein}) takes the form 
\eqn 
&& e^{\{N\}}_l =\frac 12 Tr \{ J_{h_{N+1}} \lambda_l \} = 
d\phi_N \delta_{l, N^2+1} +\frac 1{2\epsilon_N} d\theta_N Tr \left\{ 
e^{-i\phi_N \lambda_{N^2+1}} \lambda_{N^2-1} e^{i\phi_N \lambda_{N^2+1}}  
\lambda_{N^2+l} \right\} 
\cr 
&& \qquad \  +\frac 12 Tr \left\{   
e^{-i\frac {\theta_N}{\epsilon_N} \lambda_{N^2-1}} J_{h_N}  
e^{i\frac {\theta_N}{\epsilon_N} \lambda_{N^2-1}} 
e^{i\phi_N \lambda_{N^2+1}} \lambda_{N^2+l}  e^{-i\phi_N \lambda_{N^2+1}} 
\right\} \ , 
\feqn 
and using the relations in appendix \ref{app:comm} we find 
\eqn 
\underline e^{\{N\}} =\left( 
\begin{array}{ccc} 
d\phi_N & 0 & 0 \\ 
0 & \sin \phi_N \cos \phi_N d\theta_N & \sin \phi_N \cos \phi_N \frac 12 
\sum_{a=2}^N Tr \left[\frac 1{\alpha_a} J_{h_N} \lambda_{a^2-1} \right] \\ 
0 & 0 & \frac 12 \sin \phi_N 
Tr \left[ e^{-i\frac {\theta_N}{\epsilon_N} \lambda_{N^2-1}} 
J_{h_N} e^{i\frac {\theta_N}{\epsilon_N} \lambda_{N^2-1}} 
\underline M
\right]
\end{array} 
\right) \label{jj} 
\feqn 
where we introduced the $\{N\}$ index to remember that this is a $2N \times 2N$ 
matrix associated to the group $SU(N+1)$.
Here $\underline M$ is a column of matrices, $M_{2j-1} =\lambda_{j^2+1}$,
$M_{2j} =\lambda_{j^2}$, $j=1,2,\ldots, N-1$.
Formula (\ref{jj}) then reads as follows: $J_{h_N}$ is a 1-form with components
$J_{h_N,c}$, $c=1,2,\ldots, 2n-2$, with respects to the coordinates $X^c$,
defined as $X^{2j-1} =\theta_j$, $X^{2j}=\phi_j$. To find the component $(r,c)$
of (\ref{jj}) one must then take the $c-$th component of $J_{h_N,c}$
and the $r-$th component of $\underline M$ before to compute the trace.
\\
The invariant measure is then 
\eqn 
det \underline e^{\{N\}} = 
d\phi_N d\theta_N \cos \theta_N \sin^{2N-1} \phi_N 
\det 
\left( 
\frac 12 Tr \left[ e^{-i\frac {\theta_N}{\epsilon_N} \lambda_{N^2-1}} 
J_{h_N} e^{i\frac {\theta_N}{\epsilon_N} \lambda_{N^2-1}} 
\underline M
\right]
\right) \ .   \label{det} 
\feqn 
We now use the recurrence relation 
\eqn 
&& J_{h_{N}} = \lambda_{(N-1)^2+1} d\phi_{N-1} +\frac 1{\epsilon_{N-1}} 
e^{-i\phi_{N-1} \lambda_{(N-1)^2+1}} \lambda_{(N-1)^2-1} e^{i\phi_{N-1} 
\lambda_{(N-1)^2+1}} 
d\theta_{N-1} \cr 
&& \qquad \ +e^{-i\phi_{N-1} \lambda_{(N-1)^2+1}} 
e^{-i\frac {\theta_{N-1}}{\epsilon_{N-1}} \lambda_{(N-1)^2-1}} 
J_{h_{N-1}} 
e^{i\frac {\theta_{N-1}}{\epsilon_{N-1}} \lambda_{(N-1)^2-1}} 
e^{i\phi_{N-1} \lambda_{(N-1)^2+1}} \ . \label{recurrence} 
\feqn 
Computing the traces different cases arise depending on whether $j=N-1$ 
or $j < N-1$; using again the relations in appendix \ref{app:comm} it is not 
too difficult to show that the last determinant is equal to 
\eqn 
&& \det \left( 
\begin{array}{cc} 
d\phi_{N-1} \cos (N\theta_N)  & -\frac 12 \sin (N\theta_N)  
\sin (2\phi_{N-1}) d\theta_{N-1} \\ 
d\phi_{N-1} \sin (N\theta_N)  & \frac 12 \cos (N\theta_N) 
\sin (2\phi_{N-1}) d\theta_{N-1} 
\end{array} 
\right) \times \cr 
&& \times \det 
\left( 
\frac 12 \cos \phi_{N-1} Tr \left[ e^{-i\frac {\theta_{N-1}}{\epsilon_{N-1}} 
\lambda_{(N-1)^2-1}} 
J_{h_{N-1}} e^{i\frac {\theta_{N-1}}{\epsilon_{N-1}} \lambda_{(N-1)^2-1}} 
\underline M
\right]
\right) \ , \nonumber 
\feqn 
which put into (\ref{det}) in turn yields the recurrence relation 
\eqn 
det \underline e^{\{N\}} =d\phi_N d\theta_N \frac 
{\sin^{2N-1} \phi_N }{\tan^{2N-4} \phi_{N-1}}  det \underline e^{\{N-1\}} \ . 
\feqn 
which can be solved to give 
\eqn 
det \underline e^{\{N\}} =2d\theta_N d\phi_N \cos \phi_N \sin^{2N-1}  
\phi_N \prod_{a=1}^{N-1} \left[ \sin \phi_a 
\cos^{2a-1} \phi_a  d\theta_a d\phi_a \right] \ . \label{measure} 
\feqn 
This is the same result as found in \cite{Tilma}. 
\subsection{The range of the parameters} 
 
At this point we are able to determine the range of the parameters in such a 
way as to cover the whole group. We will do this only for the base space: 
The remaining ranges for the fiber can be determined recursively, as 
discussed above, remembering in particular that the $U(1)$ phase in $U(k)$ 
can be taken in $[0, 2\pi/k]$.\\ 
We then proceed as in \cite{Cacciatori:2005yb}. We first choose the ranges so 
as to generate a closed $((N+1)^2-1)-$dimensional closed manifold which then 
has to wrap around the group manifold of $SU(N+1)$ an integer number 
of times. This can be done by looking at the measure (\ref{measure}) 
on the base manifold and noticing that it is non singular 
when $0< \phi_a <\frac \pi2$, whereas $\theta_a$ can take all the period values 
$\theta_a \in [0, 2\pi]$, for all $a=1, \ldots, N$. However, note that the 
angles $\theta_1, \phi_1, \theta_2$ generate the whole $SU(2)$ group when 
$0 \leq \theta_1 \leq \pi$, $0< \phi_a <\frac \pi2$ and 
$0 \leq \theta_2 \leq 2\pi$. 
We can then restrict $\theta_1 \in [0,\pi]$. The rest of the variety is 
generated by the remaining $U(N)$ part.\\ 
If we call $V_{N+1}$ the manifold obtained this way we then find 
{\small 
\eqn 
&& Vol(V_{N+1}/U(N))  =\int_0^\pi d\theta \prod_{a=2}^N \int_0^{2\pi} d\theta_a 
\prod_{b=1}^N \int_0^{\frac \pi2} d\phi_b \left\{  
\cos \phi_N \sin^{2N-1} \phi_N \prod_{c=1}^{N-1} \left[ \sin \phi_c 
\cos^{2c-1} \phi_c  \right] \right\} \cr 
&& \qquad \qquad \qquad \qquad =\frac {\pi^N}{N!} \ , 
\feqn 
} 
or equivalently 
\eqn 
Vol (V_{N+1}) =Vol (U(N)) \frac {\pi^N}{N!} \ . 
\feqn 
This is exactly the recursion relation found in App. \ref{approot}. Therefore, 
it is the correct range of the parameters for every $N\geq 2$, if we have 
$V_3 =SU(3)$, as can be easily checked directly or by comparison with  
the results given in appendix A of \cite{Cvetic:2001zx}.\footnote{See  
also App B of \cite{Cacciatori:2005gi}.}\\ 
The next step is to determine the parametrization of $SU(N+1)$ for every 
value of $N$ . It is given by (\ref{ansatz}) with 
\eqn 
&& 0\leq \theta_1 \leq \pi \ , \qquad 0\leq \theta_a  \leq 2\pi \ , a=2, 
\ldots, N \cr 
&& 0\leq \omega \leq \frac {2\pi}N \ , \qquad 0\leq \phi_a \leq \frac \pi2 \ , 
a=1, \ldots, N \ ,  \label{range} 
\feqn 
and the remaining parameters which cover $SU(N)$ (determined inductively).\\ 
To prove that our parametrization is well-defined we can do more: We are in 
fact able to show that the induced metric on the base manifold is exactly the 
Fubini-Study metric over $\mathbb{CP}^N$. 
 
\section{The geometric analysis of the fibration} 
\label{last} 
 
We will now show that the metric induced on the base space takes exactly the 
form of the Fubini-Study metric in trigonometric coordinates as given in 
appendix \ref{appFS}. To do so we will again use inductive arguments.\\ 
The metric on the base is $ds^2_B =[\underline e^{\{ N \}}]^T \otimes  
\underline e^{\{ N \}}$, where $T$ indicates transposition and 
$\underline e^{\{ N \}}$ is given in (\ref{jj}). Using the relations in 
appendix \ref{app:comm} and defining 
\eqn 
X_N =\frac 12 \sum_{a=2}^N Tr \left[ J_{h_N} \epsilon_a \lambda_{a^2-1}  
\right] \label{XN} 
\feqn 
the metric takes the form 
\eqn 
&& ds^2_B = d^2\phi_N  +\sin^2 \phi_N \left\{ 
\left[ d\theta_N +X_N \right]^2 
+\sum_{j=1}^{N-1} \left[ \frac 12 Tr \left( e^{-i\frac {\theta_N}{\epsilon_N} 
\lambda_{N^2-1}} 
J_{h_N} e^{i\frac {\theta_N}{\epsilon_N}\lambda_{N^2-1}} \right) \lambda_{j^2} 
\right]^2 \right. \cr 
&& \qquad \ \left. +\sum_{j=1}^{N-1} \left[ \frac 12 Tr \left(  
e^{-i\frac {\theta_N}{\epsilon_N}\lambda_{N^2-1}} J_{h_N}  
e^{i\frac {\theta_N}{\epsilon_N}\lambda_{N^2-1}} \right) \lambda_{j^2 +1}  
\right]^2 \right\} -\sin^4 \phi_N \left[ d\theta_N +X_N \right]^2 \ . 
\feqn 
This is an encouraging form, which upon comparison with (\ref{cpmetric})  
suggests the identification $\xi=\phi_N$.\\ 
With this identification in mind, let's first remark that the following 
recursion relation holds 
\eqn 
X_N =\cos^2 \phi_{N-1} ( d\theta_{N-1} +X_{N-1}) \ , \label{xnn-1} 
\feqn 
which can be shown by inserting (\ref{recurrence}) in (\ref{XN}) and then 
applying (\ref{2}) and (\ref{questa}). 
A direct computation yields 
\eqn 
X_3 =\cos^2 \phi_2 (d\theta_2 +\cos (2\phi_1 ) d\theta_1 ) \ , 
\feqn 
from which, through repeated application of the recurrence relation 
(\ref{xnn-1}), we obtain 
\eqn 
X_N =\sum_{k=1}^{N-3} \left[ \prod_{i=1}^k \cos^2 \phi_{N-i} \right] 
d\theta_{N-k} +\left[ \prod_{i=1}^{N-2} \cos^2 \phi_{N-i} \right] 
(d\theta_2 +\cos (2\phi_1 ) d\theta_1 ) \ . 
\feqn 
At this point we have to compare $d\theta_N +X_N$ with the coefficient 
of $\sin^4 \xi$ in (\ref{cpmetric}). 
In fact, to bring $d\theta_N +X_N$ to the desired form 
$\sum_{i=1}^N (\tilde R^i)^2 d \psi_i$, one is tempted to just set 
$\theta_i = \psi_i $ and $\phi_\mu =\omega_\mu$. 
However, this cannot be the case because the $\tilde R^i$ don't satisfy the 
condition $\sum (\tilde R^i)^2 =1$.\\ 
These observations, together with explicit calculations for the case  
$N=4$ and $N=5$, suggest that we should simply take some linear combination 
$\psi_i =\psi_i (\theta_j)$. This can be done as follows: 
Let us introduce new variables $\tilde \theta_K \ , k=1,\ldots , N$, such that  
\eqn 
&& \tilde \theta_N =\theta_N \ , \qquad \theta_{N-k} =\tilde \theta_{N-k} 
-\tilde \theta_{N-k+1} \ , k=1 ,\ldots, N-3 \ , \cr 
&& \theta_1 +\theta_2 = \tilde \theta_1 -\tilde \theta_3 \ , \qquad \theta_1 
-\theta_2 = \tilde \theta_3 -\tilde \theta_2 \ . \label{change} 
\feqn 
In this way $d\theta_N +X_N$ takes the desired form 
\eqn 
d\theta_N +X_N =\sum_{i=1}^N (R^i (\omega_\mu ))^2 d \psi_i  
\feqn 
with $\omega_\mu =\phi_{\mu}$, $\mu= 1,\ldots N-1$, $\psi_i  
=\tilde \theta_{N-i+1}$, $i=1 , \ldots, N$ and 
\eqn 
&& R_1 =\sin \phi_{N-1} \ , \qquad R_k =\sin \phi_{N-k} \prod_{i=1}^{k-1}  
\cos \phi_{N-i} \ , k=2 ,\ldots, N-1 \ , \cr 
&& R_N =\prod_{i=1}^{N-1} \cos \phi_{N-i} \ .  \label{erre} 
\feqn  
These formulas agree with the expressions in App.~\ref{appFS}. 
As the last step, in App~\ref{appagree} we finally show that, after performing 
the change of variables described above, the coefficients of  
$\sin^2 \xi$ and $\sin^2 \phi_N$ also agree. 
This proves that the metric induced on the base $\mathbb{CP}^N$ of the $U(N)$ 
fibration by the invariant metric on $SU(N+1)$ is nothing  
else but the natural Fubini-Study metric in trigonometric coordinates.\\ 
We can now use this result as a different method to fix the range of the  
parameters. 
In fact, $(R^1, \ldots, R^N)$ parametrize the positive orthant of a sphere,  
if $0 < \phi_i <\frac \pi2 $, $i=1, \ldots, N-1$. Moreover, the identification 
of $\phi_N$ with $\xi$ yields $\phi\in [0,\pi/2]$. Finally, it is easy to show 
that the conditions $\tilde \theta_i \in [0, 2\pi]$ are equivalent to 
$\theta_1 \in [0, \pi]$ and $\theta_i \in [0, 2\pi]$, $i=2, \ldots, N$.\\ 
These are the same results obtained in (\ref{range}).


\section{Conclusions} 
\setcounter{equation}{0} 
In this paper, we have reconsidered the problem of constructing a 
generalized Euler parametrization for $SU(N)$. The parametrization we find 
differs slightly from the one described by Tilma and Sudarshan. 
In fact, comparing our results with the expression (18) in (\cite{Tilma}), 
it is possible to see that we have chosen $\lambda_{(k-1)^2-1}$  
instead of $\lambda_3$. Furthermore, we have computed the corresponding 
invariant measure, which turns out to coincide with the result in 
(\cite{Tilma}), despite the slight differences in the choice of the 
parametrization.\\ 
To determine the range of the parameters, we have used two distinct methods, 
both yielding the same result. To better motivate the name "Euler angles", 
we have carefully shown that the parametrization captures the Hopf fibration 
structure of the $SU(N)$ groups. In particular the change of coordinate we found to
evidentiate the fibrations, gives an explicit map between the Euler coordinates
introduced starting from the generalized Gell-Mann matrices, and the ones
introduced in \cite{klimyk} using geometrical considerations.\\
We have given a quite explicit proof of every assertion. Apart from 
corroborating the results of Tilma and Sudarshan, we think that our work 
is providing a complete toolbox of computation techniques useful in applied 
theoretical physics as well as for experimental physicists. 
 
\section*{Acknowledgments} 
 
SB and SC would like to thank G.~Berrino, Prof.~R.~Ferrari and Dr.~L.~Molinari
for helpful conversations.
BLC is supported by the INFN under grant 8930/01. 
This work was supported in part by the Director, Office of Science, Office of 
High Energy and Nuclear Physics, of the U.S. Department of Energy  
under Contract DE-AC02-05CH11231 and in part by the US Department of Energy,  
Grant No.DE-FG02-03ER41262.  
The work of SC is supported by INFN, COFIN prot. 2003023852\_008 and the  
European Commission RTN program MRTN--CT--2004--005104, in which SC is 
associated to the University of Milano--Bicocca.

\newpage 
\begin{appendix} 
 
\section{Some commutators} 
\label{app:comm} 
\setcounter{equation}{0} 
Using the explicit form of the generalized Gell-Mann matrices constructed 
with the conventions of section \ref{algebra}, we find the useful commutators 
\eqn 
&& [\lambda_{N^2+1} , \lambda_{N^2+2j} ]=-i\lambda_{j^2} \cr 
&& [\lambda_{N^2+1} , \lambda_{N^2+2j+1} ]=i\lambda_{j^2+1} \ , 
\feqn 
when $j=1, \ldots , N-1$. \\ 
Others interesting relations easy to check are 
\eqn 
&& [\lambda_{N^2+1} , \lambda_{N^2} ]=-i (N+1) \epsilon_{N+1} \lambda_{(N+1)^2-1}-i\sum_{a=2}^N \epsilon_a \lambda_{a^2-1} \ ,   \cr 
&& [\lambda_{N^2+1} , \lambda_{a^2 -1} ] =i\epsilon_a \lambda_{N^2} \ , \cr 
&& [\lambda_{N^2+1} , \lambda_{(N+1)^2 -1} ] =i(N+1)\epsilon_{N+1} \lambda_{N^2} \ , 
\feqn 
where $a=1, \ldots , N$, from which remembering that $\epsilon_k =\sqrt{\frac 2{k(k-1)}}$, one also finds 
\eqn 
[\lambda_{N^2+1} , [\lambda_{N^2+1} , \lambda_{N^2} ]] =4\lambda_{N^2}  \ . 
\feqn 
{From} the first two commutators we find the very useful relations 
\eqn 
&& e^{ix \lambda_{N^2+1}} \lambda_{j^2+1} e^{-ix \lambda_{N^2+1}} =\frac 1{\sin x} \lambda_{N^2+2j+1} 
-\frac 1{\tan x}  e^{ix \lambda_{N^2+1}} \lambda_{N^2+2j+1} e^{-ix \lambda_{N^2+1}} \ , \cr 
&& e^{ix \lambda_{N^2+1}} \lambda_{j^2} e^{-ix \lambda_{N^2+1}} =-\frac 1{\sin x} \lambda_{N^2+2j} 
+\frac 1{\tan x}  e^{ix \lambda_{N^2+1}} \lambda_{N^2+2j} e^{-ix \lambda_{N^2+1}} \ , \label{steve}  
\feqn 
when $j=1, \ldots , N-1$. \\ 
Other useful relations easy to prove using the previous relations are  
\eqn 
&& \sum_{a=2}^N \epsilon_a^2 +(N+1)^2 \epsilon_{N+1}^2 =4 \ , \label{4} \\ 
&& \sum_{a=2}^N \epsilon_a^2 +(N+1) \epsilon_{N+1}^2 =2 \ , \label{2} \\ 
&& Tr \left[ e^{-ix \lambda_{N^2+1}} \lambda_{N^2-1} e^{ix \lambda_{N^2+1}} \lambda_{N^2+I} \right] =\epsilon_N \delta_{I0} \sin (2x) \ , \\ 
&& \frac 12 Tr \left[ \lambda_{a} e^{ix \lambda_{N^2+1}} \lambda_{N^2+2i} e^{-ix \lambda_{N^2+1}} \right]  
= \delta_{a,i^2} \sin x    \ , a\leq N^2-1 \ , i=1,\ldots , N-1 \\ 
&& \frac 12 Tr \left[ \lambda_{a} e^{ix \lambda_{N^2+1}} \lambda_{N^2+2i-1} e^{-ix \lambda_{N^2+1}} \right]  
= -\delta_{a,i^2+1} \sin x    \ , a\leq N^2-1 \ , i=1,\ldots , N-1 \\ 
&& \frac 12 Tr \left[e^{ix \lambda_{N^2+1}} \lambda_{N^2} e^{-ix \lambda_{N^2+1}} \sum_{a=1}^{N^2-1} C^a \lambda_{a} \right] 
=\sin (2x) \frac 12 \sum_{b=2}^N Tr \left[ C^{b^2-1} \epsilon_{b^2-1} \right] \ ,  
\feqn 
\eqn  
&& \sum_{a=2}^N Tr \left[ A e^{ix \lambda_{(N-1)^2+1}} \lambda_{a^2-1} e^{-ix \lambda_{(N-1)^2+1}}  \right] 
=cos^2 x \sum_{a=2}^N Tr \left[ A \epsilon_a \lambda_{a^2-1}   \right] \ , \label{questa} \\ 
&& e^{ix\lambda_{N^2-1}} \lambda_{(N-1)^2} e^{-ix\lambda_{N^2-1}} = \cos (N\epsilon_N x) \lambda_{(N-1)^2} - 
\sin (N\epsilon_N x) \lambda_{(N-1)^2+1} \ , \\ 
&& e^{ix\lambda_{N^2-1}} \lambda_{(N-1)^2+1} e^{-ix\lambda_{N^2-1}} = \cos (N\epsilon_N x) \lambda_{(N-1)^2+1} + 
\sin (N\epsilon_N x) \lambda_{(N-1)^2} \ , 
\feqn 
where we used $A:= \sum_{i=1}^{(N-1)^2 -1} A^i \lambda_i$. 

 
\section{The total volume of \boldmath{$SU(k)$}} 
\setcounter{equation}{0} 
\label{approot} 
The total volume for the groups $SU(k)$ can be found following as shown by Macdonald in \cite{Burgy}. 
First remember that, in the sense of rational cohomology, $SU(k)$ is equivalent 
to the product of odd dimensional spheres 
\eqn 
SU(k+1) \sim \prod_{j=1}^{k} S^{2i+1} \ . 
\feqn 
where we chosen $k+1$ to obtain recursive relations. 
The total volume of the group is then uniquely determined when the a metric is 
established on the Lie algebra. We chosen the metric induced by the 
scalar product $(A|B) =\frac 12 Tr (AB)$, for $A,B\in su(k+1)$. In this way the 
Gell-Mann generators are orthonormal. The formula for the total volume is 
\cite{Burgy} 
\eqn 
Vol (SU(k+1))=\prod_{j=1}^{k} Vol(S^{2i+1}) \cdot Vol(\mathbb{T}_k) 
\prod_{\alpha>0} |\alpha^\vee |^2  \ , 
\feqn 
where $\alpha^\vee$ are the coroots associated to positive roots and 
$Vol(\mathbb{T}_k)$ is the volume of the torus generated by the simple coroots.\\ 
For $su(k+1)$ the simple coroots are $s_i =L_i-L_{i+1},\ i=1,\ldots, k$ where 
$L_i$ is the diagonal matrix with the only non vanishing entry $\{ L_i\}_{ii} =1$. 
After writing $s_i$ in terms of $\lambda_j$, as 
\eqn 
s_i =\sum_{a=1}^k \frac 12 Tr \{ s_i \lambda_{(a+1)^2-1} \} \lambda_{(a+1)^2-1} \ , 
\feqn 
it is easy to prove the recursive relation 
\eqn 
Vol(\mathbb{T}_k)=\sqrt {\frac {k+1}{2k}} Vol(\mathbb{T}_{k-1}) \ . 
\feqn 
{From} this we find 
\eqn 
Vol(SU(k+1)) =Vol(SU(k)) 2 \frac {\pi^{k+1}}{k!}  \sqrt {\frac {k+1}{2k}} 
\feqn 
where we used the fact that all the positive coroots have unitary length. 
If we note that the phase $e^{i\frac {\theta_k}{\epsilon_k} \lambda_{(k+1)^2-1}}$ 
generates an $U(1)$ group of volume $2\pi \sqrt {\frac {k(k+1)}2}$ and that 
$U(k)=\frac {SU(k)\times U(1)}{\mathbb{Z}_k}$, we can finally write 
\eqn 
Vol(SU(k+1)) =Vol(U(k))  \frac {\pi^{k}}{k!} \ . 
\feqn

\section{The Fubini-Study metric for \boldmath{$\mathbb{CP}^N$}} 
\label{appFS} 
\setcounter{equation}{0} 
$\mathbb{CP}^N$ is a K\"ahler manifold of complex dimension $N$. In a local 
chart, which uses holomorphic inhomogeneous coordinates 
$\{ z^i \}_{i=1}^N \in \mathbb{C}$, the K\"ahler potential is  
$K (z^i , \bar z^j ) =\frac k2 \log (1+\sum_{i=1}^N |z^i|^2)$ 
with $k$ a constant. The associated K\"ahler metric 
$g_{i\bar j} =\frac {\partial^2K}{\partial z^i \partial \bar z^j}$ is then 
\eqn 
ds^2_{\mathbb{CP}^N}=k \left(  
\frac {\sum_{i=1}^N dz^i d\bar z^i}{1+\sum_{i=1}^N |z^i|^2} 
-\frac {\sum_{i,j=1}^N z^i d\bar z^i \bar z^j dz^j }{(1+\sum_{i=1}^N  
|z^i|^2)^2} \right) \ . 
\feqn 
Notice that obviously it is not possible to cover the whole space with  
a single chart, but the set of points which cannot be covered has vanishing 
measure. For our purpose it is therefore enough to consider a single chart.\\ 
Let us now search for a trigonometric coordinatization. To this aim let us 
introduce the new real coordinates $\xi, \omega_\mu , \psi_i$, 
$\mu=1,\ldots,N-1$, $i =1,\ldots, N$, such that 
\eqn 
z^i =\tan \xi R^i (\omega_\mu) e^{i\psi_i} \ . 
\feqn 
Here $R^i (\omega_\mu)$ is a parametrization of the unit sphere $S^{n-1}$, 
construced as an immersion in $\mathbb{R}^N$, where $\sum_{i=1}^N (R^i)^2 =1$ 
and $\omega_\mu$ are the angles of the sphere. However, notice that we are 
restricted to the positive orthant only: $R_i >0$. If $\omega_\mu$ are the 
standard angles (starting for example with the azimuthal one $\omega_1$), 
then $\omega_\mu \in [0,\pi/2]$, $\xi \in [0, \pi/2]$ and  
$\psi_i \in [0,2\pi]$. This choice of coordinates finally gives 
\eqn 
ds^2_{\mathbb{CP}^N} =d\xi^2 +\sin^2 \xi \left[ \sum_{i=1}^N dR^i dR^i 
+\sum_{i=1}^N (R^i)^2 d^2 \psi_i  \right] -\sin^4 \xi \left[  
\sum_{i=1}^N (R^i)^2 d \psi_i \right]^2 \ .   \label{cpmetric} 
\feqn 
In particular notice that the coefficient of $\sin^2 \xi$ yields a metric for 
(the positive orthant of) the sphere $S^{N-1}$. 
 
\section{Final checks} 
\label{appagree} 
Here we verify that the change of variables introduced in section \ref{last} 
transforms the terms 
\eqn 
&& 
\left[ d\theta_N +X_N \right]^2 
+\sum_{j=1}^{N-1} \left[ \frac 12 Tr \left( e^{-i\frac {\theta_N}{\epsilon_N} 
\lambda_{N^2-1}} 
J_{h_N} e^{i\frac {\theta_N}{\epsilon_N}\lambda_{N^2-1}} \right) \lambda_{j^2} 
\right]^2 
\cr 
&& \qquad \  
+\sum_{j=1}^{N-1} \left[ \frac 12 Tr \left( e^{-i\frac {\theta_N}{\epsilon_N} 
\lambda_{N^2-1}} 
J_{h_N} e^{i\frac {\theta_N}{\epsilon_N}\lambda_{N^2-1}} \right) 
\lambda_{j^2 +1} \right]^2 \ , 
\label{check} 
\feqn 
into the coefficient of $\sin^2 \xi$ in (\ref{cpmetric}).\\ 
First, using (\ref{recurrence}) and the relations in appendix $\ref{app:comm}$, 
it is possible to show that 
\eqn 
&& Tr \left\{ e^{-i\frac {\theta_N}{\epsilon_N} \lambda_{N^2-1}} J_{h_N} 
e^{i\frac {\theta_N}{\epsilon_N} \lambda_{N^2-1}} \lambda_{j^2} \right\}  \cr 
&& \qquad =\cos \phi_{N-1}  
Tr \left\{ e^{-i\frac {\theta_{N-1}}{\epsilon_{N-1}} \lambda_{(N-1)^2-1}} 
J_{h_{N-1}} 
e^{i\frac {\theta_{N-1}}{\epsilon_{N-1}} \lambda_{(N^2-1)-1}} \lambda_{j^2} 
\right\} \ , \  j< N-1 
\cr 
&& Tr \left\{ e^{-i\frac {\theta_N}{\epsilon_N} \lambda_{N^2-1}} J_{h_N} 
e^{i\frac {\theta_N}{\epsilon_N} \lambda_{N^2-1}} \lambda_{j^2+1} \right\}  \cr 
&& \qquad =\cos \phi_{N-1}  
Tr \left\{ e^{-i\frac {\theta_{N-1}}{\epsilon_{N-1}} \lambda_{(N-1)^2-1}} 
J_{h_{N-1}} 
e^{i\frac {\theta_{N-1}}{\epsilon_{N-1}} \lambda_{(N-1)^2-1}} \lambda_{j^2+1} 
\right\} \ , \  j< N-1 
\cr 
&& Tr \left\{ e^{-i\frac {\theta_N}{\epsilon_N} \lambda_{N^2-1}} J_{h_N} 
e^{i\frac {\theta_N}{\epsilon_N} \lambda_{N^2-1}} \lambda_{(N-1)^2} \right\}\cr 
&& \qquad  =\sin (2\phi_{N-1}) \cos (N\theta_N) \left[ d\theta_{N-1} + 
X_{N-1} \right] -2\sin (N\theta_N) d\phi_{N-1} \ , \cr 
&& Tr \left\{ e^{-i\frac {\theta_N}{\epsilon_N} \lambda_{N^2-1}} J_{h_N} 
e^{i\frac {\theta_N}{\epsilon_N} \lambda_{N^2-1}} \lambda_{(N-1)^2+1} \right\} 
\cr 
&& \qquad  =\sin (2\phi_{N-1}) \sin (N\theta_N) \left[ d\theta_{N-1} + 
X_{N-1} \right] +2\cos (N\theta_N) d\phi_{N-1} \ . 
\feqn 
Note that these are true for $N\geq 3$, if we define  
$X_2 :=cos (2\phi_1 )d\theta_1$. 
{From} these relations we find 
\eqn 
&& Tr \left\{ e^{-i\frac {\theta_N}{\epsilon_N} \lambda_{N^2-1}} J_{h_N} 
e^{i\frac {\theta_N}{\epsilon_N} \lambda_{N^2-1}} \lambda_{j^2} \right\}  \cr 
&& \qquad =\left[ \prod_{k=j+1}^{N-1} \cos \phi_k \right] 
\left[ \sin (2\phi_j )\cos \left[ (j+1)\theta_{j+1}\right] (d\theta_j +X_j)  
-2\sin \left[(j+1)\theta_{j+1} \right] d\phi_j \right] \ , \cr 
&& Tr \left\{ e^{-i\frac {\theta_N}{\epsilon_N} \lambda_{N^2-1}} J_{h_N} 
e^{i\frac {\theta_N}{\epsilon_N} \lambda_{N^2-1}} \lambda_{j^2+1} \right\}  \cr 
&& \qquad =\left[ \prod_{k=j+1}^{N-1} \cos \phi_k \right] 
\left[ \sin (2\phi_j )\sin \left[(j+1)\theta_{j+1}\right] (d\theta_j +X_j) 
+2\cos \left[(j+1)\theta_{j+1}\right] 
d\phi_j \right] \ , \nonumber   
\feqn 
with $j=2, \ldots, N-1$. For $j=1$ 
\eqn 
&& Tr \left\{ e^{-i\frac {\theta_N}{\epsilon_N} \lambda_{N^2-1}} J_{h_N} 
e^{i\frac {\theta_N}{\epsilon_N} \lambda_{N^2-1}} \lambda_{1} \right\}  \cr 
&& \qquad =\left[ \prod_{k=2}^{N-1} \cos \phi_k \right] 
\left[ \sin (2\phi_1 )\cos (2\theta_2) d\theta_1 -\sin (2\theta_2 ) 
d\phi_1 \right] \cr 
&& Tr \left\{ e^{-i\frac {\theta_N}{\epsilon_N} \lambda_{N^2-1}} J_{h_N} 
e^{i\frac {\theta_N}{\epsilon_N} \lambda_{N^2-1}} \lambda_2 \right\}  \cr 
&& \qquad =\left[ \prod_{k=2}^{N-1} \cos \phi_k \right] 
\left[ \sin (2\phi_1 )\sin (2\theta_2 ) d\theta_1 +\cos (2\theta_2) 
d\phi_1 \right] \ . \nonumber  
\feqn 
Thus we see that (\ref{check}) takes the form $S_N+U_N$, where 
\eqn 
&& S_N =d\phi^2_{N-1} +\sum_{j=1}^{N-2} \left[ \prod_{k=j+1}^{N-1} 
\cos^2 \phi_k \right] d\phi^2_j \ , \label{esse} \\ 
&& U_N =(d\theta_N+X_N)^2 +\sum_{j=2}^{N-1}\sin^2 \phi_j 
\left[ \prod_{k=j}^{N-1} \cos^2 \phi_k \right] (d\theta_{j} +X_j )^2 \cr 
&& \qquad \ +\prod_{k=2}^{N-1} \cos^2 \phi_k \sin^2 (2\phi_1) d\theta^2_1 \ .  
\label{u} 
\feqn 
First, we show that 
\eqn 
S_N =\sum_{j=1}^N dR^j dR^j \ , 
\feqn 
with $R^j$ as in (\ref{erre}). To this aim let us define the  
$N$-dimensional vector $\vec R_N =(R^1,\ldots,R^N)$. Such a vector has unit  
length, and satisfies the recurrence relation 
$\vec R_N =(\sin \phi_{N-1} , \cos \phi_{N-1} \vec R_{N-1})$, from which we 
find 
\eqn 
d\vec R_N \cdot d\vec R_N =d\phi^2_{N-1} +\cos^2 \phi_{N-1} d\vec R_{N-1} 
\cdot d\vec R_{N-1} \ . 
\feqn 
Here the dot indicates the scalar product in $N$ dimensions. Now, from 
(\ref{esse}), we also have 
\eqn 
S_N =d\phi^2_{N-1} +\cos^2 \phi_{N-1} S_{N-1} \ . 
\feqn 
As $S_N$ and $d\vec R_N \cdot d\vec R_N$ both satisfy the same recurrence 
relation, the thesis follows because of $S_2=d\vec R_2 \cdot d\vec R_2$.\\ 
The second and last step of our proof consists in showing that after the change 
of coordinates (\ref{change}) the equation (\ref{u}) takes the form 
\eqn 
U_N =\sum_{i=1}^N (R^i)^2 d\psi_i^2 \ . 
\feqn 
The structure of (\ref{u}) suggests that it is convenient to make the  
change of variables starting from $\theta_N$ and $\theta_{N-1}$ step by step. 
Note that $X_{N-1}$ is invariant under this transformation, so that we have 
\eqn 
&&(d\theta_N +X_N)^2 +\sin^2 \phi_{N-1} \cos^2 \phi_{N-1} (d\theta_{N-1}  
+X_{N-1})^2 \cr && 
\qquad \qquad \qquad \qquad \qquad \qquad \qquad =\sin^2 \phi_{N-1}  
d \tilde \theta^2_N +\cos^2 \phi_{N-1} (d\tilde \theta_{N-1} 
+X_{N-1})^2 \ .  \label{tnn-1} 
\feqn 
Here we have used (\ref{xnn-1}) to express $X_N$ in terms of $X_{N-1}$. 
Then $U_N$ takes the form 
\eqn 
&& U_N=\sin^2 \phi_{N-1} d \tilde \theta^2_N +\cos^2 \phi_{N-1} 
\left[(d\tilde \theta_{N-1} +X_{N-1})^2 \right. \cr 
&& \qquad \ \left. +\sin^2 \phi_{N-2} \cos^2 \phi_{N-2} 
(d\theta_{N-2} +X_{N-2})^2  \right] +\ldots \ . 
\feqn 
Now it is possible to use (\ref{tnn-1}) with $N-1$ in place of $N$ in order 
to write $\theta_{N-2}$ in terms of $\tilde \theta_{N-2}$. In fact, this 
relation can be applied recursively up to $d\theta_3$, obtaining 
\eqn 
&& U_N = \sin^2 \phi_{N-1} d\tilde \theta^2_N 
+\sum_{j=2}^{N-4} \sin^2 \phi_{N-j} \left[ \prod_{l=N-j+1}^{N-1} \cos^2 
\phi_l \right] d\tilde \theta^2_{N-j+1} \cr 
&& \qquad \ +\left[\prod_{l=3}^{N-1} \cos^2 \phi_l \right] (d \tilde 
\theta_3 +X_3)^2 
+\sin^2 \phi_2 \left[\prod_{k=2}^{N-1} \cos^2 \phi_k \right] (d\theta_2  
+\cos (2\phi_1) 
d\theta_1)^2 \cr 
&& \qquad \ +\left[ \prod_{k=2}^{N-1} \cos^2 \phi_k \right] \sin^2 (2\phi_1) 
\theta^2_1 \ . 
\feqn 
At this point we can perform the last two changes of coordinates in 
(\ref{change}), to show that 
\eqn 
d \tilde \theta_3 +X_3 =\sin^2 \phi_2 d \tilde \theta_3 + 
\cos^2 \phi_2 ( \sin^2 \phi_1 \tilde \theta_2 +\cos^2 \phi_1 d \tilde 
\theta_1 ) \ , 
\feqn 
and this completes the proof. 
 
 
\end{appendix}

\end{document}